\documentclass[aps,prb,twocolumn,showpacs,preprintnumbers,amsmath,amssymb,superscriptaddress]{revtex4-1}

\usepackage{graphicx}
\usepackage{dcolumn}
\usepackage{color}
\usepackage{bm}
\usepackage{amsfonts}
\usepackage{array}
\usepackage{txfonts}

\newcommand{\nc}{\newcommand}
\nc{\be}{\begin{equation}}
\nc{\ee}{\end{equation}}
\nc{\bea}{\begin{eqnarray}}
\nc{\eea}{\end{eqnarray}}
\nc{\bean}{\begin{eqnarray*}}
\nc{\eean}{\end{eqnarray*}}
\nc{\mb}{\mbox}
\nc{\rnc}{\renewcommand}
\nc{\vk}{\mb{\bf k}}
\nc{\vp}{\mb{\bf p}}
\nc{\vn}{\mb{\bf n}}
\nc{\vq}{\mb{\bf q}}
\nc{\rr}{\mb{\bf r}}
\nc{\vz}{\hat {\mb{\bf z}}}
\nc{\vj}{\mb{\boldmath$j$}}
\nc{\vg}{\mb{\boldmath$g$}}
\nc{\x}{\mb{\boldmath$x$}}
\nc{\A}{\mb{\boldmath$A$}}
\nc{\va}{\mb{\boldmath$a$}}
\nc{\vs}{\mb{\boldmath$\sigma$}}
\nc{\vpi}{\mb{\boldmath$\pi$}}
\nc{\nab}{\nabla}
\nc{\X}{\sf x}

\begin{document}

\title{Adiabatic cooling of Majorana zero-modes in topological superconductors}

\author{Chang-Yu Hou}
\thanks{These two authors contributed equally}
\affiliation{Department of Physics and Astronomy, University of California
at Riverside, Riverside, CA 92521}
\affiliation{Department of Physics, California Institute of Technology,
Pasadena, CA 91125}

\author{Yafis Barlas}
\thanks{These two authors contributed equally}
\affiliation{Department of Physics and Astronomy, University of California
at Riverside, Riverside, CA 92521}

\author{Kirill Shtengel}
\affiliation{Department of Physics and Astronomy, University of California
at Riverside, Riverside, CA 92521}
\affiliation{Institute for Quantum Information, California Institute of Technology,
Pasadena, CA 91125}

\date{\today}

\begin{abstract}
We show that the presence of Majorana zero-modes in a one-dimensional topological superconductor can be detected by adiabatic cooling. This cooling effect results from an increase of the topological entropy
associated with the ground state degeneracy due to such modes. Here, we consider
an experimentally feasible topological superconductor: a strong spin-orbit
coupled semiconducting quantum wire interfaced with an $s$-wave
superconductor and subjected to a magnetic field. Numerical simulations of
realistic experimental geometries indicate that the creation of Majorana
zero-modes, which can be achieved by tuning of electronic gates or the
external magnetic field, results in a measurable cooling effect. We also
argue that this cooling effect results in an increase in the zero bias peak
conductance.
\end{abstract}

\maketitle

{\em Introduction}---Majorana zero-modes, which are predicted to appear in
odd-pairing topological superconductors,~\cite{Read2000,Kitaev2001} have
recently attracted a lot of attention. This interest is partially driven by
the possibility of employing non-Abelian statistics associated with the
Majorana zero-modes for topological quantum
computation.~\cite{Kitaev2003,Nayak2008} It has been suggested that
topological superconductors can be realized in heterostructure devices built
from readily available ingredients, such as topological
insulators~\cite{Fu2008} or Rashba spin-orbit coupled
semiconductor,~\cite{Sau2010,Alicea2010} in proximity with an $s$-wave
superconductor. Among all proposals, devices that consists of a
one-dimensional semiconducting wires with strong spin-orbit coupling in
proximity with an ordinary $s$-wave superconductor are the most
promising.~\cite{Lutchyn2010,Oreg2010} This optimism is fueled by the recent
experimental observation of a zero-bias peak, which is likely due to the
presence of Majorana zero-modes.~\cite{Law2009,Mourik2012,Das2012,Deng2012}
The appearance of this zero-bias peak, however, does not provide any
information about the quantum statistics associated with the Majorana
zero-modes. In this paper, we propose an experimental procedure to verify the
Majorana-zero mode quantum statistics by detecting their topological entropy.

Majorana-zero modes have proven challenging to detect in two-dimensional
settings. In quantum Hall systems, experimental proposals are often based on
interferometry,~\cite{DasSarma2005,Stern2006,Bonderson2006} which requires a
microscopic understanding of edge transport, or charge sensitive bulk
measurements,~\cite{Yang2009,Cooper2009,Barlas2012} neither of which applies
to a 1D topological superconductor. Alternatively, a cooling effect
associated with the ground state degeneracy of Majorana zero-modes, similar
to spin demagnetization cooling effect, has been proposed for charge
insensitive bulk detection.~\cite{Gervais2010,Yamamoto2011} As a necessary
condition for non-Abelian statistics, the ground state degeneracy, $\Gamma
\sim d^{N_{q}}$, has to grow exponentially with respect to the number of
quasiparticles $N_q$, where $d>1$ is the quantum dimension of the non-Abelian
quasiparticles ($d = \sqrt{2}$ for Majorana zero-modes). This contributes a
{\em temperature independent} topological entropy,~\cite{footnote1} $S_{D} =
k_{\rm B} N_{q} \ln d + O(1)$, to the total entropy,
\begin{equation}
\label{eq:S_tot}
S(T) = S_{D}+ S_{n}(T),
\end{equation}
where $S_{n}(T)$ is the normal entropy that
monotonically increases as a function of the temperature. When Majorana
zero-modes are created in an adiabatic process, i.e., $S(T)$ remains
constant, the increase of the topological entropy leads to a reduction of the
normal entropy and hence a cooling effect.

\begin{figure}
\begin{center}
\includegraphics[angle=0,scale=1]{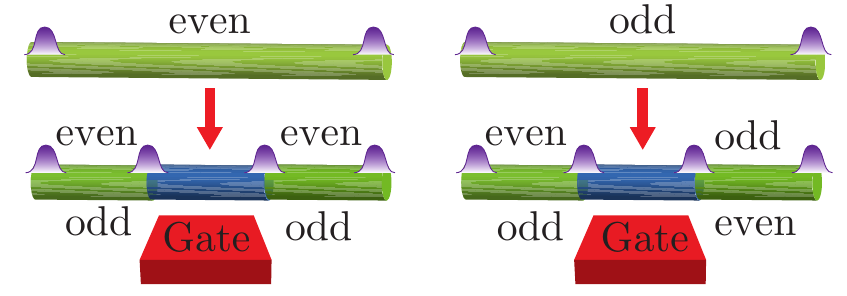}
\caption{Schematic representation of an even and odd parity topological superconductor (shown in green). As the middle section of the wire is tuned to the non-topological phase (shown in blue) Majorana zero-modes are generated at the domain walls. For each case, two possible configurations for the occupation of the non-local fermion formed by a pair of Majorana zero modes give the topological entropy $S_{D} = k_{\rm B} \ln 2$ associated with the creation of Majorana zero modes.}
\label{fig:schematic}
\end{center}
\end{figure}

In this paper, we consider the adiabatic cooling effect due to the creation
of Majorana zero-modes in a one-dimensional topological superconductor made
of a semiconducting quantum wire in proximity with an $s$-wave
superconductor. We start with a wire in the topological superconducting phase
by  properly tuning of the chemical potential and the external magnetic
field.~\cite{Alicea2011} Gating and variation of the magnetic field is used
to bring the middle segment of the wire into a non-topological phase as shown
in Fig~\ref{fig:schematic}. As a consequence, additional localized Majorana
modes appear at the domain wall separating the topological and
non-topological phases. When this procedure is performed adiabatically, it
results in a cooling effect, i.e., the final temperature is lower than the
initial temperature $T_f < T_i$. Based on a numerical analysis of
experimentally accessible setups with realistic material parameters, we show
that an adiabatic cooling effect of $\Delta T \sim 20$ mK can be achieved. As
the cooling effect leads to an increase in the zero bias conductance peak, it
can be detected in transport.

{\em Entropy expression for topological superconductors}---Topological superconductors have a degeneracy resulting from the fact that a pair of Majorana zero-modes form a non-local fermion with zero-energy. The occupied fermionic state has the ground state with odd parity, whereas the unoccupied fermion state has even parity. Now, the contributions to the topological entropy can simply be understood by counting the possible ground states configurations. Consider a topological superconductor with a fixed parity,~\cite{footnote2} tuning the middle segment of the wire to the non-topological phase results in only two choices. For example, if we begin with an even parity superconductor, the outer segments of topological superconductors can have either even or odd parity after tuning as shown in Fig.~\ref{fig:schematic}. Thus the excess entropy associated with this procedure is just $k_{\rm B} \ln 2 = 2 k_{\rm B} \ln \sqrt{2}$ (exactly the entropy corresponding to two Majorana zero-modes). A similar counting argument holds for an odd parity superconductor.

This simple counting argument can be formally derived as follows. Since the total entropy of a superconductor with fixed particle parity is different from one where particle parity can fluctuate, we shall start from the partition function with the parity constraint~\cite{Janko1994}
\begin{equation}
\label{eq:Z-mean-field}
Z^{e(o)}\equiv {\rm Tr}\left[ \frac{1\pm(-1)^N}{2} e^{-\beta H_{\rm MF}}\right] = \frac{1}{2} \left( \mathcal{Z}_+ \pm \mathcal{Z}_{-} \right),
\end{equation}
where the superscript $e(o)$ corresponds to even(odd) parity and $N$ is the total particle number. The projection operator $[(1 \pm (-1)^N)/2]$ suppresses contributions from states with odd and even particle parity, respectively. The mean field Hamiltonian, $H_{\rm MF} = \sum_{n} E_n \gamma_n^{\dag} \gamma_n$, is expressed in terms of the Bogoliubov quasiparticle operators $\gamma_{n}$ with quasiparticle energy $E_n\geq 0$. In terms of the energy spectrum, we have $\mathcal{Z}_{\pm} =\prod (1 \pm e^{-\beta E_n})$. In making the BCS approximation, we have assumed that the low-temperature dynamics are completely governed by independent Bogoliubov quasiparticles.

The total entropy can be evaluated from the partition function as
\begin{equation}
S^{e(o)} =-k_{\rm B} {\rm Tr} \left[\rho^{e(o)} \ln \rho^{e(o)} \right],
\end{equation}
where $\rho_{e(o)}$ is the density matrix for even(odd) parity, is given by
\begin{equation}
\rho^{e(o)}= \frac{\left(1 \pm (-1)^N \right) e^{-\beta H_{\rm MF}} }{\mathcal{Z}_+ \pm \mathcal{Z}_{-}}.
\end{equation}
Using Eq.~\eqref{eq:Z-mean-field}, the total entropy can be expressed as
\begin{equation}
\label{eq:entropy-e/o}
\frac{S^{e(o)}}{k_{\rm B}} =  \sum_{n} \beta E_n \left( \frac{f_+ \mathcal{Z}_+ \pm f_- \mathcal{Z}_-}{\mathcal{Z}_+ \pm \mathcal{Z}_-} \right)  + \ln {Z^{e(o)}} ,
\end{equation}
where $f_{\pm}(E_n)= 1/ (1\pm e^{\beta E_n})$. We note that the term inside the parentheses of Eq.~\eqref{eq:entropy-e/o} is the average occupation number $\langle n(E_n)\rangle_{e(o)}$ of a quasiparticle state with the energy $E_n$ under the even(odd) parity constraint.

In the presence of $m \ge 1$ fermionic zero-energy states,
Eq.~\eqref{eq:entropy-e/o} can be simplified to
\begin{equation}
\label{eq:entropy-m-zero-modes}
\frac{S(T)}{k_{\rm B}} = (m-1) \ln 2 + \left(\sum_{n} \beta E_n  f_+(E_n)  + \ln \mathcal{Z}'_+ \right) ,
\end{equation}
where we have defined $\mathcal{Z}'_+ \equiv \mathcal{Z}_+/2^{m}$ to allow
for separation of the zero-energy state contributions. Now,
Eq.~\eqref{eq:entropy-m-zero-modes} represent all the contributions to the
entropy of a superconducting system with a parity constraint in the presence
of fermionic zero-energy mode(s). Identifying a pair of Majorana zero-modes
with a fermionic zero-energy state, i.e., $N_q=2m$, facilitates comparison of
Eq.~\eqref{eq:entropy-m-zero-modes} with the general entropy expression of a
non-Abelian system (see Eq.~\eqref{eq:S_tot}). The first term in
Eq.~\eqref{eq:entropy-m-zero-modes} can be recognized as the temperature
independent topological entropy $S_D$ of $N_{q}$ Majorana zero-modes, while
the terms inside the parentheses represent the temperature dependent entropy
due to the normal sources $S_n(T)$

{\em Cooling effect of quantum wire} --- As mentioned in the introduction, gating or variation of the magnetic field will generate additional Majorana zero-modes, which will lead to a cooling effect. For this cooling effect to be observable, it is important that (i) the normal sources provide enough initial entropy, which requires an initial minimum temperature $T > T_{min}$ for the experiment, and (ii) the normal entropy contribution should be strongly temperature dependent and weakly dependent on other parameters. Here, the minimum initial temperature, defined by $S_{n}(T_{min}) = k_{\rm B} \ln 2$, is the lowest temperature required to generate additional Majorana zero-modes adiabatically. The second condition requires that the quasiparticle density of states in the energy window, $E \lesssim T_i$, does not change significantly in the cooling process, where the energy $E$ is measured respect to the superconductor ground state energy.

For a finite superconducting quantum wire, the fermionic contributions to the entropy come from either discrete energy states, due to size confinement, or a continuous Bogoliubov spectrum $E > \Delta$ ($\Delta$ is the proximity induced superconducting gap). The continuous Bogoliubov density of states changes drastically, as the number of occupied bands increases by one. Therefore, in order to satisfy condition (ii), we must exclude the entropy contributions of the continuous Bogoliubov quasiparticle states. This can be achieved by choosing the initial temperature of experiment much smaller than the superconducting gap. In contrast, the discrete states correspond to confined Andreev bound states resulting from finite size effects and typically lie inside the superconducting gap. Since these discrete states are determined by the wire dimensions both in the topological and the non-topological phase, their density of states remains unchanged during the transition. In order to ensure that the change in the normal contribution to entropy during the tuning procedure is associated to a reduction in temperature only, we numerically check that the density of states does not change significantly for the entire parameter window considered below. Combining the condition discussed above with (i), we require the existence of the temperature regime $k_{\rm B} T_{min} < k_{\rm B} T_i \ll \Delta$ to observe the cooling effect. Below, using numerical simulation, we show that the above temperature regime exists and leads to a measurable cooling effect.

Let us begin by describing the one-dimensional semiconducting wire with strong spin-orbit coupling in proximity with an $s$-wave superconductor with a tight-binding Hamiltonian in the $N_x\times N_y \times N_z$ rectangular lattice sites:~\cite{Liu2012}
\begin{multline}
\label{eq:H_L}
H_{L} = \sum_{\boldsymbol{r}, \boldsymbol{d},\alpha, \beta} c^{\dag}_{\boldsymbol{r} +\boldsymbol{d},\alpha} \left[  - t  \delta_{\alpha \beta} -i U_R \hat{z} \cdot \left( \sigma_{\alpha \beta} \times \hat{d} \right) \right] c_{\boldsymbol{r},\beta}
\\
- \sum_{\boldsymbol{r},\alpha,\beta} c^{\dag}_{\boldsymbol{r},\alpha} \left[ \mu(\boldsymbol{r}) \delta_{\alpha \beta} + \frac{g \mu_B}{2} \boldsymbol{B}\cdot \boldsymbol{\sigma}_{\alpha\beta} \right] c_{\boldsymbol{r},\beta}
\\
+\sum_{\boldsymbol{r},\alpha} \Delta_{0}(\boldsymbol{r}) \delta_{r_{z},0} \left( c^{\dag}_{\boldsymbol{r},\alpha}  c^{\dag}_{\boldsymbol{r},-\alpha}  +
c_{\boldsymbol{r},-\alpha}  c_{\boldsymbol{r},\alpha} \right) ,
\end{multline}
where $c_{\boldsymbol{r},\alpha} (c^{\dag}_{\boldsymbol{r},\alpha})$ is the annihilation (creation) operator at the site $\boldsymbol{r} = (r_{x},r_{y},r_{z})$ with spin $\alpha=\uparrow, \downarrow $, and $\boldsymbol{d}=a \hat{d}$ is the vector connecting the nearest neighboring sites. The following parameters are chosen to coincide with the material parameters and experimental conditions: the hopping constant $t$, the lattice spacing $a$, the spin-orbit coupling strength on the lattice $U_R$, the chemical potential $\mu(\boldsymbol{r})$, and the Zeeman energy $E_z= g \mu_B |\boldsymbol{B}|/2$ with the magnetic field $\boldsymbol{B}$ ($g$ is the Land\' e g-factor and $\mu_B$ is the Bohr magneton). The superconductor contact is modeled by introducing a constant pairing potential $\Delta_{0}(\boldsymbol{r}) = \Delta_0$ at the bottom of $x-y$ surface, which self-consistently gives a desired proximity induced pairing potential $\Delta$. As both the magnetic field and the local chemical potential can be tuned externally, a segment of the wire can be driven into the topological phase when  $E_Z > \sqrt{\Delta^2 + (\mu(\boldsymbol{r})-\mu_{0,n})^2}$ is satisfied.~\cite{Alicea2011} Here, $\mu_{0,n}$ is the chemical potential corresponding to the center of the gap between the $2n-1$ and the $2n$ bands of the multi-band model in Eq.~\eqref{eq:H_L}.

As a concrete example, we focus on the InSb semiconductor wire that has the band mass $m=\hbar^2/(2 t a^2)=0.015 m_e$, spin-orbit energy $E_{\rm SO}=U_R^2/t=50$ $\mu$eV, and $g\approx50$.~\cite{Mourik2012} We set $N_x\times N_y \times N_z=600\times 5 \times 4$ with the lattice spacing $a=25$ nm that leads to the hopping amplitude $t= 4.06$~meV and $U_R \simeq 0.11 t$. Throughout the discussions, we choose the chemical potential $\mu \approx \mu_{0,3}$, to sit around the gap between the $5^{\rm th}$ and $6^{\rm th}$ bands. The choice of the pairing potential  $\Delta_0= 650$ $\mu$eV, self-consistently leads to a proximity induced pairing potential $\Delta \approx 200$ $\mu$eV,~\cite{Mourik2012} choosen to mimic experimental conditions. A magnetic field applied along the wire direction $B_x$, gives the Zeeman energy $E_z\approx 1.45 B_x$ meV/T.

We now propose two possible schemes to implement the adiabatic cooling effect
due to the creation of the Majorana zero-modes labelled as (I) and (II) from
here onwards. (I) Start with a uniform chemical potential $\mu_{i}\approx
\mu_{0,n}$ occupying $2n-1$ bands with $E_Z > \Delta$. The whole wire is in
the topological phase with a pair of Majorana zero modes at its ends. Then
adiabatically tune the chemical potential of the middle segment of the wire
to $\mu^{m}_{ f}=\mu_{i}+ d\mu^{m}$. This drives the middle segment into the
non-topological phase and creates two extra Majorana zero-modes at the domain
walls. (II) Start with a non-uniform chemical potential profile
$\mu_{i}(\boldsymbol{r})$ with a higher value in the middle segment of the
wire. A sufficiently large magnetic field is applied to the wire so that the
whole wire is initially in the topological phase. Then adiabatically reduce
the magnetic field to drive the middle segment of the wire into the
non-topological phase creating two extra Majorana zero-modes at the domain
walls.

\begin{figure}
\includegraphics[angle=0,scale=0.95]{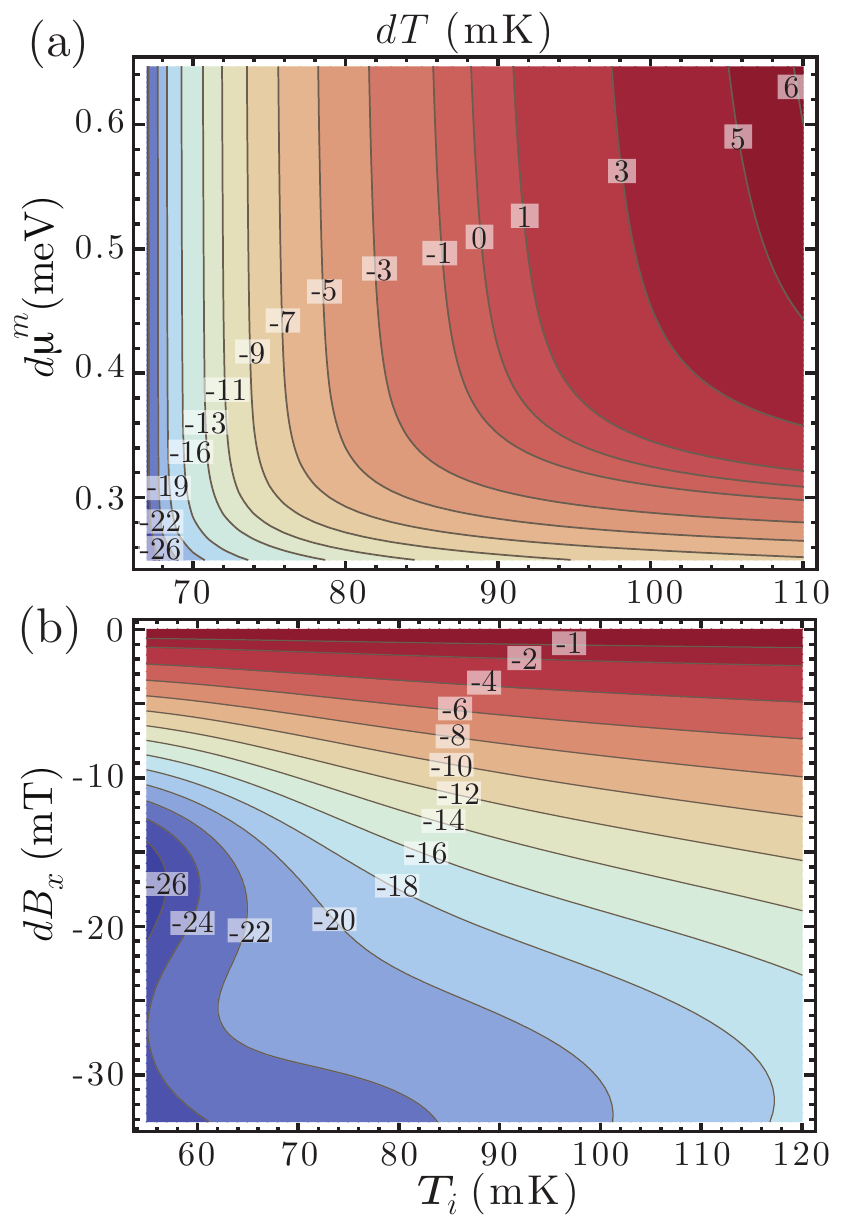}
\caption{(a) The contour plot shows the temperature difference $dT= T_{f}- T_{i}$ as the function of the initial temperature $T_{i}$ and the chemical potential difference $d\mu^m = \mu^{m}_{f}- \mu_{i}$ which tuned by the gate placed across the middle segment of the wire. Here, the initial chemical potential, $\mu_i=\mu_{0,3}$ (defined in the text), is chosen to have the whole wire in the topological phase. (b) The contour plot shows the temperature difference $dT= T_{f}- T_{i}$ as the function of the initial temperature $T_{i}$ and the magnetic field difference $d B_x = B_{x,f}- B_{x,i}$. As discussed in the text, the chemical potential of the middle segment of the wire is higher than that of two outer segments, $\mu^m> \mu^o$ while a sufficient strong magnetic field $B_{x,i}$ is initially applied to have the whole wire in the topological phase.}
\label{fig:dT-change}
\end{figure}

To demonstrate that the adiabatic cooling effect can be observed in the InSb
semiconductor wire setup, we numerically compute the spectrum of the lattice
model, Eq.~\eqref{eq:H_L} with the aforementioned material parameters. Then,
the entropy of the system can be evaluated from
Eq.~\eqref{eq:entropy-m-zero-modes}. From the initial setup of either scheme
(I) or scheme (II) at an initial temperature $T_{i}$, the total entropy is
determined by $S_{i} (T_{i})$. Here the subscript $i(f)$ denotes the
initial(final) state of the setup. By adiabatically creating an extra pair of
Majorana zero-modes following the procedure described in either scheme (I) or
scheme (II), the temperature independent topological entropy becomes
$S_D=k_{\rm B} \ln 2$ from Eq.~\eqref{eq:entropy-m-zero-modes}. As the total
entropy is unchanged in an adiabatic process, the final temperature $T_{f}$
can be evaluated by requiring that $S_{f}(T_{f})= S_D+ S_{n,f}(T_{f}) =
S_{i}(T_{i})$.

The implementation of scheme (I) gives the temperature difference $dT= T_{f}
- T_{i}$  shown in the contour plot in Fig.~\ref{fig:dT-change}a as a
function of the initial temperature $T_i$ and change of the chemical
potential $d\mu^{m}= \mu^{m}_{f}- \mu_{i}$. An external magnetic field
$B_x\approx 200$ mT is applied along the wire direction leads to the Zeeman
energy $E_Z\approx 285$ $\mu$eV. The initial choice of chemical potential
$\mu_{i} = \mu_{0,3}$, requires $d \mu^{m} \gtrsim 200$ $\mu$eV to drive the
middle segment of the wire into the non-topological phase. The cooling effect
can be realized with a low enough initial temperature $T_{i} \lesssim 88$ mK
and effectively becomes independent of $d \mu^{m}$ when the middle segment of
the wire is deep inside the non-topological phase, $d \mu^{m}\gtrsim 350$
$\mu$eV. This stems from the fact that the quasiparticle density of state
(other than the Majorana zero-modes) in the energy window, $E \lesssim T_i$,
remain largely unchanged over the gating process. It is worthwhile to note
that the initial entropy $S_{i}(T_i)< k_{\rm B} \ln 2$ when $T_i \lesssim 67$
mK which sets the minimum temperature $T_{min}$.

Following scheme (II), the contour plot in Fig.~\ref{fig:dT-change}b shows the temperature difference $dT= T_{f} - T_{i}$ as functions of the initial temperature $T_i$ and the difference of the applied magnetic field $dB_{x}= B_{x,f}- B_{x,i}$ in the $x$-direction. We use the external gates to tune the chemical potential of two outer segments and the middle segment of the wire to $\mu^{o}= \mu_{0,3}$ and $\mu^{m}= \mu_{0,3} + 165$~$\mu$eV, respectively. Initially, an external magnetic field $B_{x,i} \approx 200$~mT is applied to the wire. Then, the magnetic field is gradually reduced to a final strength $B_{x,f}$. In the current setup, the middle segment of the wire becomes non-topological, hence creating a pair or Majorana zero-modes, when the magnetic field difference $dB_x =B_{x,f}-B_{x,i} \lesssim - 18$~mT. We note that scheme (2) results in more irregular cooling effect as tuning the magnetic field in general causes more dramatic changes of quasiparticle density of state.

Provided that the hybridization~\cite{footnote1} of Majorana modes is still negligible, some comments regarding to the qualitative changes upon altering experimental parameters are in order. First, because the total entropy is an extensive quantity, a shorter wire or fewer occupied bands provide a smaller normal entropy reservoir and, in principle, has a more prominent cooling effect. Second, increasing the number of gates to create more Majorana zero-modes leads to a stronger cooling effect in a given wire. However, both changes will lead to a higher minimum temperature $T_{min}$ below which the process becomes non-adiabatic, hence the temperature regime in which adiabatic cooling process can be detected becomes smaller. In contrast to the quantum Hall case, where the temperature change is determined by the density and the magnetic field, the temperature change in our case depends on the initial conditions. However, as the operational temperature window in the quantum Hall case~\cite{Gervais2010} is of the same magnitude as in our analysis, we expect that the magnitude of the cooling effect to be similar in both cases.


{\em Discussion}---This cooling effect can be detected in transport by analyzing the response of the differential conductance $G(eV,T)$ with respect to the temperature. Here, we will focus on the temperature dependence of the zero-bias conductance peak (ZBCP) observed in experiments.~\cite{Mourik2012} Since the Majorana zero-mode obeys Fermi-Dirac distribution, the difference of the peak value, $\Delta G = G(0,T_{f}) - G(0,T_{i})$, under the cooling process, at low-temperatures, can be approximated as
\begin{equation}
\Delta G \simeq G^{''}(0,T=0) \frac{\pi^2}{3}  T_{ave} \Delta T,
\end{equation}
where $\Delta T = T_{f} -T_{i}$ is the change in temperature, $T_{ave}= (T_{f}+T_{i})/2$ and the double primes in the superscript indicates the double derivatives with respect to $eV$. If the differential conductance $G(eV,T=0)$ is maximized at $V=0$, as seen in the current experiments, the cooling will lead to an increase in the ZBCP, while heating will lead to a decrease in the ZBCP. Assuming a Lorentzian profile of the conductance peak with the broadening $\gamma \sim 25$ $\mu$eV,~\cite{Mourik2012} we estimate the variation of conductance to be $5 \sim 10 \%$ of $G_0 \sim 2 e^2/h$. We would like to stress that only a relative sign change of the temperature needs to determined to detect a cooling effect, therefore, alternate methods such as thermopower~\cite{Hou2013} or optical techniques can also be employed.

To satisfy the adiabaticity, entropy should not flow in and out of the system
as the quasiparticle density is varied throughout the cooling process. This
can be achieved by thermally disconnecting the topological superconductor
from the environment. In our system, the bulk $s$-wave superconductor in
proximity with the wire provides a reservoir that carries no entropy and
effectively isolates the wire.~\cite{Gervais2010} In addition, to minimize
entropy losses during the tuning process, we require a slow thermal exchange
rate between the lattice and the electrons, i.e., phonons need to be frozen
at low temperature, which will lead to a long enough thermal relaxation time,
$\tau_T$. Since quasiparticle poisoning will alter the parity of the
topological superconductor by introducing non-equilibrium particles, the
proposed adiabatic cooling schemes must also be performed faster than the
poisoning time scale, $\tau_{qp}$. Alternately, the electronic responsive
time scale, $\tau_s$, which is of the order of nanoseconds, is fast enough to
maintain quasi-static equilibrium in the whole system throughout the tuning
process. In order to ensure adiabaticity, the tuning procedure has to be
performed within the time scale $\tau_s< \tau <{\rm min }(\tau_T,\tau_{qp})$.
Typical superconducting qubits have $ \tau_{qp} \sim 10 \mu s - 1 m s$,
indicating that experimental procedures proposed here can be performed
adibatically.~\cite{Manucharyan2012,Riste2013}

To conclude, we would like to comment on other contribution to the normal
entropy, such as phonons and domain wall fluctuations which have been
ignored. In the low temperature regime, phonons are mostly frozen out and
their contributions are negligible. The domain wall fluctuations between the
topological and the non-topological regions can lead to extra entropy,
however, as they are pinned by the external gates, they can be treated
classically. Other sources of entropy, for example, trapped spin, 
local fermion modes e.t.c., only matter if the entropy associated to these 
modes changes during the tuning procedure, in which case complementary 
measurements are required to rule out these possibilities.
Finally, the presence of scalar potential disorder
alters the quasiparticle spectrum~\cite{Sau2013} which can change the entropy
carried by the quasiparticles. With weak disorder our numerical analysis
shows that the cooling effect is unaltered. However, the presence of strong
disorder requires more careful consideration, which is beyond the scope of
the work.

We are grateful to D.~Pekker, K.~Yang, D.~J.~Clark, R.~Mong, J.~Sau, and
M.~Wimmer for useful discussions. CYH, YB and KS were supported in part by
the DARPA-QuEST program. KS was supported in part by NSF award DMR-0748925.
The authors would also like to acknowledge support from the Aspen Center of 
Physics during the intial stages of this work.

\bibliographystyle{apsrev}
\bibliography{cooling}

\begin{thebibliography}{30}
\bibitem[{foo({\natexlab{a}})}]{footnote1}
\bibinfo{note}{Except below an energy scale which is related to the coupling
  between the Majorana zero-modes. The energy splitting of Majorana zero-modes
  decays exponentially with respect to the ratio $\ell/\xi$, where $\ell$ is
  the distance separating two Majorana modes and $\xi$ is the superconductor
  coherence length.}
\bibitem[{foo({\natexlab{b}})}]{footnote2}
\bibinfo{note}{Since the parity of the topological superconductor is fixed,
  there is no topological entropy associated with the initial even or odd
  occupation configuration.}
\end{thebibliography}

\end{document}